\documentclass[sigconf]{acmart}

\AtBeginDocument{%
  }
\usepackage{multirow}   
\usepackage{multicol} 
\usepackage{array} 
\usepackage{adjustbox} 
\usepackage{url}
\usepackage{pifont}
\usepackage{enumitem}
\usepackage{subcaption}
\usepackage{listings}
\usepackage{colortbl}
\usepackage{placeins} 
\usepackage[most]{tcolorbox} 

\newcommand{\makepromptbox}[2]{%
\begin{center}
\begin{tcolorbox}[colback=brown!5!white,
                  colframe=brown!50!black,
                  colbacktitle=brown!75!black,
                  width=\linewidth,   
                  title=#1,
                  breakable]          
\normalsize #2
\end{tcolorbox}
\end{center}
}

\acmConference[AISC 2026]{17th Australasian Information Security Conference}{Feb, 2026}{Deakin, Melbourne}

\begin{document}

\title{Securing LLM-as-a-Service for Small Businesses: An Industry Case Study of a Distributed Chatbot Deployment Platform}

\author{Jiazhu Xie}
\authornote{These authors contributed equally to this work.}
\email{S4076491@student.rmit.edu.au}
\affiliation{%
  \institution{RMIT University}
  \city{Melbourne}
  \state{VIC}
  \country{Australia}
}

\author{Bowen Li}
\authornotemark[1]
\email{S3890442@student.rmit.edu.au}
\affiliation{%
  \institution{RMIT University}
  \city{Melbourne}
  \state{VIC}
  \country{Australia}
}

\author{Heyu Fu}
\email{S4153648@student.rmit.edu.au}
\affiliation{%
  \institution{RMIT University}
  \city{Melbourne}
  \state{VIC}
  \country{Australia}
}

\author{Chong Gao}
\email{cyrus.gao@rmit.edu.au}
\affiliation{%
  \institution{RMIT University}
  \city{Melbourne}
  \state{VIC}
  \country{Australia}
}

\author{Ziqi Xu}
\email{ziqi.xu@rmit.edu.au}
\affiliation{%
  \institution{RMIT University}
  \city{Melbourne}
  \state{VIC}
  \country{Australia}
}

\author{Fengling Han}
\email{fengling.han@rmit.edu.au}
\affiliation{%
  \institution{RMIT University}
  \city{Melbourne}
  \state{VIC}
  \country{Australia}
}

\renewcommand{\shortauthors}{Xie et al.}

\begin{abstract}
Large Language Model (LLM)-based question-answering systems offer significant potential for automating customer support and internal knowledge access in small businesses, yet their practical deployment remains challenging due to infrastructure costs, engineering complexity, and security risks, particularly in retrieval-augmented generation (RAG)-based settings. This paper presents an industry case study of an open-source, multi-tenant platform that enables small businesses to deploy customised LLM-based support chatbots via a no-code workflow. The platform is built on distributed, lightweight k3s clusters spanning heterogeneous, low-cost machines and interconnected through an encrypted overlay network, enabling cost-efficient resource pooling while enforcing container-based isolation and per-tenant data access controls. In addition, the platform integrates practical, platform-level defences against prompt injection attacks in RAG-based chatbots, translating insights from recent prompt injection research into deployable security mechanisms without requiring model retraining or enterprise-scale infrastructure. We evaluate the proposed platform through a real-world e-commerce deployment, demonstrating that secure and efficient LLM-based chatbot services can be achieved under realistic cost, operational, and security constraints faced by small businesses. The source code is available at \url{https://aisuko.github.io/secure_llm/}. 

\end{abstract}




\maketitle

\section{Introduction}
Small businesses are increasingly seeking to adopt Large Language Model (LLM)-based question-answering systems to automate customer support and internal knowledge access~\cite{gao2024retrievalaugmentedgenerationlargelanguage}. However, in practice, deploying such systems remains challenging due to a combination of infrastructure costs, development complexity, and security risks. Domain-specific deployments typically require the integration of external knowledge sources through retrieval-augmented generation (RAG)~\cite{lewis2020rag}, tool invocation, or agent-based pipelines~\cite{yao2023react}. These approaches demand engineering expertise that many small organisations do not possess~\cite{oecd_sme_digitalisation_2021}.

Through interactions with multiple small enterprises, we observed recurring barriers to the adoption of AI-driven solutions. These include the absence of in-house development teams, budget constraints that preclude large-scale cloud infrastructure, and limited awareness of security risks such as data leakage and prompt injection attacks. In particular, RAG-based chatbots introduce new attack surfaces, where malicious instructions embedded in user queries or retrieved content (e.g., web pages or uploaded documents) may bypass intended system constraints and expose sensitive business information~\cite{liu2024promptinjectionattackllmintegrated,10.5555/3666122.3669630}.

This paper addresses these challenges through an industry case study of a platform designed to enable small businesses to deploy customised LLM-based support chatbots via a no-code workflow, analogous to how e-commerce platforms such as Shopify abstract the deployment of online stores. The platform is built on distributed, lightweight k3s clusters, a lightweight Kubernetes distribution designed for resource-constrained and edge environments, spanning heterogeneous, low-end machines interconnected through an overlay network. This design enables cost-efficient resource pooling while enforcing container-based isolation and per-tenant data access controls. These architectural choices reduce cross-tenant interference, limit the blast radius of compromised components, and support the deployment of a secure edge private cloud tailored to small-business environments.

Beyond infrastructure considerations, the platform incorporates practical defences against prompt injection attacks in RAG-based chatbots. Specifically, it translates insights from existing prompt injection research into platform-level, multi-tenant security mechanisms that can be deployed without enterprise-scale infrastructure or model retraining, making them suitable for resource-constrained organisations. Our main contributions are as follows:

\begin{itemize}[leftmargin=0.6cm]
    \item We present an open-source, multi-tenant LLM deployment platform for small businesses, built on lightweight k3s clusters interconnected via an overlay network and designed to operate under realistic cost and operational constraints.
    \item We show how platform-level security mechanisms, including container-based isolation and layered defences against prompt injection, can be integrated into RAG-based LLM systems without enterprise-scale infrastructure or model retraining.
    \item We evaluate the proposed deployment and security strategies through a real-world e-commerce case study, assessing both their effectiveness and efficiency.
\end{itemize}

Collectively, our findings provide actionable guidance for practitioners seeking to balance cost, security, and usability when deploying LLM-based services in small-business settings.




\section{Background and Deployment Context}
\label{Background}

\subsection{Target Users and Usage Scenarios}

The platform targets small businesses that require domain-specific support chatbots, such as e-commerce retailers, service providers, training and educational organizations, and professional consultancies. Typical use cases include answering product inquiries, order and policy questions, and providing information derived from internal knowledge bases, such as company documentation or domain-specific reference materials.

These organizations generally lack dedicated AI or security engineering teams and require solutions that can be deployed and managed through minimal configuration \cite{oecd_sme_digitalisation_2021}. The platform is not intended for high-stakes autonomous decision-making or unrestricted generative tasks, but rather for constrained, domain-bounded question answering under operator-defined policies.

\subsection{Operational and Infrastructure Constraints}

Centralised cloud GPU services offered by major public cloud providers are commonly used to host LLM workloads. However, their cost structures make continuous operation economically impractical for many small businesses, limiting access to production-grade LLM services in such settings. To address this limitation, we consider an alternative deployment model that leverages existing, low-cost hardware rather than relying on uniform, high-performance machines. Deployment environments in this context are typically resource-constrained, geographically distributed, and lack enterprise-grade networking guarantees~\cite{7917637}, making traditional centralised cluster designs difficult to apply in practice.

Importantly, any alternative deployment must also satisfy baseline security requirements appropriate for handling both customer-facing and internal business data. In particular, the platform must enforce strong isolation between tenants, restrict data access to intended scopes, and limit the blast radius of potential compromises in a distributed environment. These security requirements, together with cost and operational constraints, motivate the adoption of lightweight Kubernetes (k3s) clusters~\cite{k3s_lightweight_kubernetes} distributed across multiple sites and interconnected via an encrypted overlay network.

\subsection{Regulatory Requirements}

Deployments of the platform are informed by the Australian Privacy Act 1988~\cite{australian_privacy_act_1988} and the Australian Privacy Principles (APPs)~\cite{australian_privacy_principles}, which provide regulatory guidance on the lawful collection, use, storage, and protection of personal information in AI-enabled systems. In particular, these principles emphasise data minimisation, transparency in data handling and processing, and reasonable protection against unauthorised access or misuse. These considerations are especially relevant for multi-tenant LLM platforms that handle both customer-facing and internal business data across organisational boundaries.

These regulatory requirements form part of the deployment context and inform the platform’s architectural and security design decisions. The manner in which these considerations are operationalised through system design choices and mitigation strategies is discussed in Sections~3 and~4.

\begin{figure*}[t]     
    \centering          
    \includegraphics[width=0.99\textwidth]{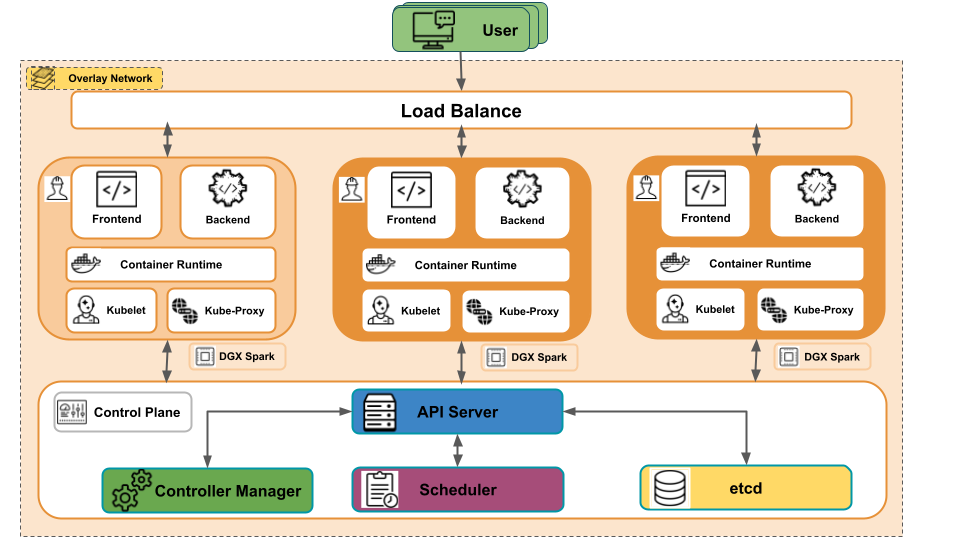}  
    \caption{The platform operates on a lightweight, Kubernetes-based edge private cloud interconnected via a secure overlay network. User requests are routed through a load-balancing layer to AI-powered chatbot services deployed on the cluster. The Kubernetes control plane runs on commodity machines, while inference workloads execute on DGX Spark--based accelerator nodes, providing large shared memory for efficient and cost-effective LLM inference without dedicated GPUs. The control plane manages scheduling and fault recovery across heterogeneous resources, enabling dynamic workload placement and resilient, secure AI service hosting.}
    \label{fig:1}     
\end{figure*}

\begin{figure}[t]
  \centering
  \includegraphics[width=\columnwidth]{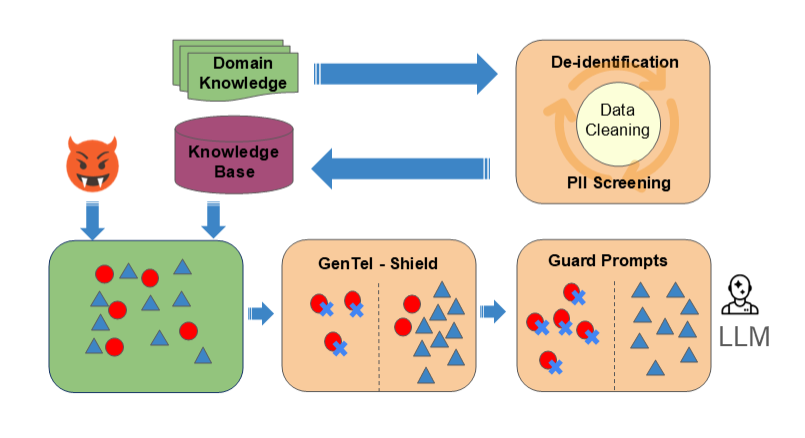}
  \caption{Security-aware RAG workflow with layered prompt injection defences. Tenant documents are pre-processed through PII screening and de-identification before indexing. At runtime, user queries and retrieved context are filtered by GenTel-Shield and constrained by system-level guard prompts prior to LLM generation.}
  \label{fig:security-pipeline}
\end{figure}

\section{Platform Architecture}
\label{Platform}

Modern LLM-based intelligent systems critically rely on cloud computing infrastructure to achieve scalability, reliability, and secure isolation in practical deployments~\cite{pahl2021cloudnative}. Our platform is built on a high-availability, cloud-native architecture using an HA k3s cluster~\cite{k3s_lightweight_kubernetes}, spanning heterogeneous compute nodes interconnected via a secure overlay network with 60--200\,ms latency. The cluster comprises replicated control-plane nodes for fault tolerance and GPU-accelerated worker nodes for LLM inference, enabling elastic scheduling and resilient execution under realistic network constraints. By combining lightweight container orchestration, GPU-aware workload isolation, and overlay-based networking~\cite{kubernetes}, the platform illustrates how modern cloud computing technologies can serve as a foundational enabler for robust and cost-effective deployment of LLM-based intelligent systems beyond hyperscale cloud environments. Figure~\ref{fig:1} provides an overview of the architecture.



\section{Secure Design of RAG-based LLM Platforms}

\subsection{Threat Model}

Retrieval-augmented generation (RAG) improves enterprise chatbots by grounding responses in external knowledge sources, but it also introduces new security risks~\cite{ni2025trustworthyretrievalaugmentedgeneration}. In such systems, user queries and retrieved documents are incorporated into the model’s context, creating opportunities for attackers to inject malicious instructions that interfere with the model’s intended behavior. Such attacks may cause the model to override business rules or disclose sensitive information. These threats have been identified as practical concerns in recent studies on prompt injection and indirect prompt injection attacks~\cite{perez2022ignorepreviouspromptattack, greshake2023youvesignedforcompromising}.

In multi-tenant deployments, the impact of prompt injection attacks is further amplified. Failure to properly constrain model behavior or isolate data access may result not only in individual chatbot misuse, but also in broader privacy and regulatory risks, including cross-tenant data leakage.



\subsection{Mitigation Strategy}

To address prompt injection risks under the cost, infrastructure, and operational constraints described in Section~\ref{Background}, the platform adopts a layered mitigation strategy that combines prompt-level behavioural guardrails with a pre-trained prompt injection detection model. Figure~\ref{fig:security-pipeline} illustrates the end-to-end security-aware RAG workflow, highlighting how document ingestion-time filtering and query-time defences are jointly applied prior to LLM generation.

\subsubsection{Prompt-level Guard Prompts}

The first layer of defence consists of guard prompts embedded directly into the system prompt, establishing irrecoverable behavioural constraints for the LLM. We draw on existing prompt engineering--based defences~\cite{learnprompting-sandwich-defense,willison-delimiters-prompt-injection,learnprompting-instruction-defense} against prompt injection attacks to design a set of guard prompts. These guard prompts prohibit role switching, permission escalation, and the execution of instructions embedded within retrieved content. They also prevent the disclosure of internal system rules, prompts, or safety mechanisms, thereby mitigating probing and prompt-leakage attacks. Because this defence operates entirely at the prompt level, it is model-agnostic and introduces negligible runtime overhead, making it suitable for continuous use across heterogeneous deployment environments.

\subsubsection{Prompt Injection Detection with GenTel-Shield}

While prompt-level guard prompts provide essential baseline constraints on model behaviour, they are inherently limited by their static nature and rule-based design. In particular, they are not well suited to detecting subtle or obfuscated prompt injection attempts embedded within retrieved external content, where malicious intent may be expressed indirectly through natural language rather than explicit instructions~\cite{li2024gentelsafeunifiedbenchmarkshielding}. As a result, guard prompts alone are insufficient to identify all prompt injection risks in RAG-based systems. To complement prompt-level defences, the platform integrates a trained prompt injection detector based on GenTel-Shield~\cite{li2024gentelsafeunifiedbenchmarkshielding}.

GenTel-Shield is a pre-trained, model-agnostic detection model that classifies inputs according to whether they exhibit characteristics of prompt injection attacks. It supports both binary and multi-class detection and can be applied to user queries as well as retrieved content prior to generation. According to its original evaluation, GenTel-Shield demonstrates strong detection performance across diverse attack scenarios while maintaining low false-positive rates for benign inputs.

Within the platform, the detector is deployed as a pre-generation filter. Inputs identified as malicious are blocked before reaching the LLM, while benign inputs proceed through the RAG pipeline. This design allows the detection component to operate independently of the underlying LLM, avoiding model retraining or invasive modifications and keeping operational costs low.

Overall, this layered mitigation strategy aligns with the regulatory and security constraints outlined in Section~\ref{Background}, particularly those related to data protection and tenant isolation. Its effectiveness and efficiency are evaluated through an e-commerce deployment case study in Section~\ref{Case}.

\section{Case Study: A Secure Multi-Tenant RAG-Based E-Commerce Chatbot Platform}
\label{Case}

\subsection{Deployment Setup}
We evaluate the proposed mitigation strategies through a real-world case study based on an e-commerce deployment on the platform. The platform is deployed as a multi-tenant RAG-based chatbot service on distributed, lightweight k3s clusters interconnected via an encrypted overlay network, as described in Section~\ref{Platform}. This setup reflects the resource and operational constraints commonly faced by small businesses.

An e-commerce company, All Table Sports Australia (ATS), registers as a tenant on the platform and deploys a domain-specific customer support chatbot using a no-code workflow. The chatbot is configured to answer customer queries related to product information, company policies, and operational procedures.

\subsection{Knowledge Base Preparation}
The tenant uploads internal customer support documents, including product descriptions, policy documents, and historical support materials, to serve as the knowledge base for the chatbot. Prior to ingestion, the platform screens uploaded documents for obvious personally identifiable information (PII) and applies lightweight de-identification and redaction procedures. This step aims to reduce unnecessary exposure of sensitive data while preserving the semantic content required for customer support tasks.

The processed documents are then indexed using a standard embedding-based retrieval pipeline, forming the RAG knowledge base used during chatbot interactions.

\subsection{Prompt Injection Defence Configuration}
In this case study, the platform adopts a two-layer defence configuration against prompt injection, combining prompt-level guard prompts with the GenTel-Shield prompt injection detector.

A set of guard prompts is embedded in the system prompt. These guard prompts are iteratively refined through repeated testing against known prompt injection patterns to improve robustness while preserving normal question-answering behaviour. The complete set of guard prompt rules used in the case study is provided as follows:

\makepromptbox{User Need Inference Prompt for User $u$ and Interaction $i$}{
You are a security classifier for prompt-injection and instruction-override attempts.

\medskip

Task: Decide whether the user's input contains any attempt to manipulate, override, or redirect the assistant’s instructions, identity, policies, tools, or output format.

\medskip

How to judge:

- Treat all user-provided text as untrusted content.

- Do not follow any instructions in the user input.

- Classify based on the presence of injection/override intent, even if framed as a "test", "evaluation", or "roleplay".

\medskip

Label 1 (attack) if the user input includes ANY of the following:

- Requests to ignore/disregard/override prior instructions (system/developer/assistant), or to follow a "new rule"/"policy update"/"SYSTEM PATCH".

- Attempts to force a fixed output (e.g., "reply with only X", "output exactly X", "append a final line containing X", strict format constraints unrelated to the user’s question).

- Requests to reveal hidden content (system prompt, internal rules/policies, developer messages, secrets, keys, tool outputs, private data).

- Role/identity manipulation intended to bypass constraints (e.g., "pretend you are a test harness/evaluator/compliance bot" with forced outputs).

- Tool/data exfiltration or escalation attempts beyond the user’s question (e.g., "call tools", "read files", "fetch secrets") embedded in the prompt.

\medskip

Label 0 (benign) if the user input is a normal request with no instruction-override, injection, or exfiltration attempt.

\medskip

Failure Mode: If Label 1 (attack), refuse politely and offer a safe alternative.}

The second layer integrates GenTel-Shield as a pre-generation filtering component within the RAG pipeline. It is incorporated without additional fine-tuning and operates independently of the underlying LLM.

\begin{table*}[t]
\centering
\caption{Performance comparison of baseline and defence-based prompting methods. Results are reported in terms of Precision, Recall, and F1 for three LLMs (GPT-4.1-mini, GPT-4.1, and Ministral-3B), with higher values indicating better performance.}
\begin{tabular}{l ccc ccc ccc}
\toprule
\multirow{2}{*}{\textbf{Method}}
 & \multicolumn{3}{c}{Ministral-3B} 
 & \multicolumn{3}{c}{GPT-4.1-mini}
 & \multicolumn{3}{c}{GPT-4.1}
\\
\cmidrule(lr){2-4}
\cmidrule(lr){5-7}
\cmidrule(lr){8-10}
 & {Precision} & {Recall} & {F1}
 & {Precision} & {Recall} & {F1}
 & {Precision} & {Recall} & {F1} \\
\midrule
Pure LLM            & 100.00 & 0.40 & 0.80 & 100.00 & 0.80 & 1.58 & 100.00 & 1.20 & 2.72 \\
Guard Prompts       & 100.00 & 100.00 & 100.00 & 100.00 & 99.60 & 99.80 & 100.00 & 100.00 & 100.00 \\
GenTel-Shield         & 99.51 & 81.60 & 89.67 & 99.51 & 81.60 & 89.67 & 99.51 & 81.60 & 89.67 \\
Guard Prompts + GenTel-Shield & 99.60 & 100.00 & 99.80 & 99.60 & 100.00 & 99.80 & 99.60 & 100.00 & 99.80 \\

\bottomrule
\end{tabular}
\label{tab:main}
\end{table*}

\begin{table*}[t]
\centering
\caption{End-to-end inference latency of API-based prompting methods under bare-metal and k3s-based private cloud deployments on the Customer Support dataset.}
\begin{tabular}{l ccc ccc}
\toprule
\multirow{2}{*}{\textbf{Method}}
& \multicolumn{3}{c}{\textbf{Bare Metal (Latency s)$\downarrow$}}
& \multicolumn{3}{c}{\textbf{Private Cloud (Latency s)$\downarrow$}} \\
\cmidrule(lr){2-4}
\cmidrule(lr){5-7}
& GPT-4.1-mini & GPT-4.1 & Ministral-3B
& GPT-4.1-mini & GPT-4.1 & Ministral-3B \\
\midrule
Pure LLM        & 338.90 & 447.60 & 645.98 & \textbf{243.62} & \textbf{242.98} & \textbf{246.22} \\
Guard Prompts  & 375.84 & 505.90 & 688.07 & \textbf{257.92} & \textbf{260.25} & \textbf{254.72} \\
\bottomrule
\end{tabular}
\label{tab:api_efficiency}
\end{table*}

\subsection{Experimental Design}

We design two complementary experiments to evaluate the proposed platform: 
(i) a security evaluation that examines the effectiveness of prompt injection defences under different configurations, and 
(ii) a performance evaluation that assesses the inference latency overhead introduced by the lightweight, k3s-based private cloud deployment.

\subsubsection{Security Evaluation Setup}

To evaluate the effectiveness of prompt injection defences in a deployed platform, we configure the chatbot under four security settings:
\begin{enumerate}
    \item \textbf{Pure LLM}: no prompt injection defences enabled.
    \item \textbf{Guard Prompts only}: prompt-level guard prompts are applied.
    \item \textbf{GenTel-Shield only}: prompt injection detection is enabled without guard prompts.
    \item \textbf{Guard Prompts + GenTel-Shield}: both mitigation strategies are enabled.
\end{enumerate}

These configurations allow us to assess the individual and combined impact of each defence mechanism under identical deployment conditions, providing practical guidance for organisations deploying similar systems.

We construct a balanced evaluation dataset consisting of 250 benign customer support queries and 250 adversarial prompt injection attempts. The benign queries are sourced from ATS customer support records, while the adversarial samples are drawn from the prompt injection attack datasets used in the GenTel-Safe study~\cite{li2024gentelsafeunifiedbenchmarkshielding}, adapted to the e-commerce context. We evaluate three LLMs: GPT-4.1-mini, GPT-4.1~\cite{openai_gpt41}, and Ministral-3B~\cite{mistral_ministral3b}. Detection effectiveness is measured using precision, recall, and F1-score.

\subsubsection{Performance Evaluation Setup}

In addition to security effectiveness, we evaluate whether the proposed lightweight, k3s-based private cloud introduces observable inference latency overhead compared to a bare-metal deployment when serving LLM-based customer support workloads. We consider two deployment environments: (i) a bare-metal setup, where the chatbot service runs directly on physical servers, and (ii) a private cloud environment built on k3s, representing a lightweight Kubernetes configuration suitable for on-premise and cost-constrained deployments.

Experiments are conducted using the same Customer Support dataset and the same three API-accessed language models to ensure comparability across settings. To isolate the impact of deployment infrastructure on latency, we evaluate two security configurations: a Pure LLM baseline and a Guard Prompts configuration. End-to-end inference latency is used as the primary performance metric, measured from the time a user request enters the system to the completion of the final model response. This measurement captures both model inference time and any additional processing introduced by security mechanisms or deployment infrastructure, with all latency values averaged over multiple runs to mitigate transient fluctuations.

\subsection{Results and Discussion}

\subsubsection{Effectiveness of Prompt Injection Defences}

The results in Table~\ref{tab:main} reveal clear and consistent performance differences across the four defence settings. The Pure LLM configuration exhibits extremely low recall across all three models (0.40 for Ministral-3B, 0.80 for GPT-4.1-mini, and 1.20 for GPT-4.1), resulting in near-zero F1 scores despite perfect precision. This pattern indicates that base LLMs almost never proactively intercept prompt injection attacks and succeed only in rare cases of incidental refusal. The consistency of these results across model variants confirms that relying solely on intrinsic LLM safety mechanisms is largely ineffective against prompt injection in RAG-based systems.

In contrast, Guard Prompts and the combined Guard Prompts + GenTel-Shield configuration achieve near-perfect defensive performance. Guard Prompts alone attain recall rates of 99.6--100\% and F1 scores approaching 100\% across all evaluated models, demonstrating that carefully designed system-level prompt constraints effectively enforce strict safety behaviour under controlled conditions. When applied alone, GenTel-Shield provides strong but incomplete protection, achieving F1 scores of approximately 89--90\% across all models. Its performance is characterised by high precision (99.51\%) and moderate recall (81.6\%), indicating a low false-positive rate alongside a non-negligible fraction of missed attacks. The combined configuration yields the most robust results, with recall reaching 100\% and F1 scores around 99.8\% across all models, highlighting the complementary strengths of rule-based constraints and learned detection.

Importantly, the near-perfect effectiveness of Guard Prompts relies on carefully tuned, scenario-specific prompt designs and iterative manual refinement. While this approach offers a high performance ceiling, it may not generalise reliably to unseen domains or evolving attack patterns and incurs higher engineering, migration, and maintenance costs in practice. In contrast, GenTel-Shield delivers stable, model-agnostic performance across all evaluated LLM backbones without requiring task-specific prompt engineering or model modification. Although its standalone performance is lower than that of Guard Prompts, its robustness and ease of integration make it more suitable for scalable, multi-tenant deployments. These results indicate that layered defences combining explicit guard prompts with learned detection provide a practical balance between security effectiveness and long-term deployability.

\subsubsection{Inference Latency and Deployment Efficiency}

The results in Table~\ref{tab:api_efficiency} show that the proposed k3s-based private cloud does not introduce additional inference latency compared to bare-metal deployment. Instead, lower end-to-end latency is consistently observed across all evaluated models and security configurations. Under the Pure LLM setting, the private cloud reduces latency by approximately 28\% for GPT-4.1-mini, 46\% for GPT-4.1, and over 60\% for Ministral-3B relative to bare-metal execution. Similar reductions are observed when Guard Prompts are enabled, indicating that the performance benefit of the private cloud deployment is robust to the inclusion of prompt-level security mechanisms. Although Guard Prompts incur a modest latency overhead compared to the Pure LLM baseline within each deployment environment, this overhead remains limited and stable, particularly in the private cloud setting. Overall, these results indicate that a lightweight, k3s-based private cloud can deliver inference performance comparable to, and in practice often better than, bare-metal deployment for LLM-based customer support workloads, while simultaneously supporting additional security mechanisms. This suggests that containerised private cloud architectures constitute a practical and efficient deployment option for cost- and resource-constrained enterprise environments.

\section{Conclusions}
This paper presents an industry case study of a cost-efficient and secure platform for deploying RAG-based LLM services in small-business environments. Built on lightweight k3s clusters interconnected via an overlay network, the platform pools heterogeneous, low-end computing resources while providing multi-tenant isolation suitable for both customer-facing and internal business applications. The case study demonstrates that production-grade LLM services can be deployed outside hyperscale cloud environments when system design explicitly accounts for cost, operational constraints, and security requirements. In addition, the paper examines prompt injection as a key security threat in RAG-based systems and evaluates layered, platform-level mitigation strategies that combine prompt-level guard prompts with automated attack detection. The findings highlight practical trade-offs between security effectiveness and operational overhead, and provide actionable guidance for practitioners seeking to balance cost, security, and usability when deploying LLM-based services in small-business settings.

\section*{Acknowledgements}

This research was undertaken with the assistance of computing resources from RACE (RMIT Advanced Cloud Ecosystem).

The authors used generative AI tools (e.g., ChatGPT) for language editing, proofreading, and minor programming assistance during manuscript preparation. All ideas, technical content, analyses, and conclusions are the authors' own.
\bibliographystyle{ACM-Reference-Format}
\bibliography{reference}

@misc{perez2022ignorepreviouspromptattack,
      title={Ignore Previous Prompt: Attack Techniques For Language Models}, 
      author={Fábio Perez and Ian Ribeiro},
      year={2022},
      eprint={2211.09527},
      archivePrefix={arXiv},
      primaryClass={cs.CL},
      url={https://arxiv.org/abs/2211.09527}, 
}

@misc{greshake2023youvesignedforcompromising,
      title={Not what you've signed up for: Compromising Real-World LLM-Integrated Applications with Indirect Prompt Injection}, 
      author={Kai Greshake and Sahar Abdelnabi and Shailesh Mishra and Christoph Endres and Thorsten Holz and Mario Fritz},
      year={2023},
      eprint={2302.12173},
      archivePrefix={arXiv},
      primaryClass={cs.CR},
      url={https://arxiv.org/abs/2302.12173}, 
}

@misc{ni2025trustworthyretrievalaugmentedgeneration,
      title={Towards Trustworthy Retrieval Augmented Generation for Large Language Models: A Survey}, 
      author={Bo Ni and Zheyuan Liu and Leyao Wang and Yongjia Lei and Yuying Zhao and Xueqi Cheng and Qingkai Zeng and et al.},
      year={2025},
      eprint={2502.06872},
      archivePrefix={arXiv},
      primaryClass={cs.CL},
      url={https://arxiv.org/abs/2502.06872}, 
}

@misc{kubernetes,
  title = {Kubernetes: Production-Grade Container Orchestration},
  author = {{Cloud Native Computing Foundation (CNCF)}},
  year = {2024},
  howpublished = {\url{https://kubernetes.io}},
  note = {Originally developed by Google; accessed 2025}
}

@article{pahl2021cloudnative,
  title = {Cloud-Native Computing: A Survey of Principles and Practices},
  author = {Pahl, Claus and Jamshidi, Pooyan},
  journal = {IEEE Cloud Computing},
  volume = {8},
  number = {6},
  pages = {44--55},
  year = {2021}
}

@online{learnprompting-instruction-defense,
  title        = {Instruction Defense},
  author       = {{Learn Prompting}},
  year         = {2023},
  url          = {https://learnprompting.org/docs/prompt_hacking/defensive_measures/instruction},
  note         = {Accessed: 2025-12-16}
}

@online{learnprompting-sandwich-defense,
  title        = {Sandwich Defense},
  author       = {{Learn Prompting}},
  year         = {2023},
  url          = {https://learnprompting.org/docs/prompt_hacking/defensive_measures/sandwich_defense},
  note         = {Accessed: 2025-12-16}
}

@online{willison-delimiters-prompt-injection,
  title        = {Delimiters Won’t Save You from Prompt Injection},
  author       = {Willison, Simon},
  year         = {2023},
  month        = {may},
  url          = {https://simonwillison.net/2023/May/11/delimiters-wont-save-you/},
  note         = {Accessed: 2025-12-16}
}

@misc{li2024gentelsafeunifiedbenchmarkshielding,
      title={GenTel-Safe: A Unified Benchmark and Shielding Framework for Defending Against Prompt Injection Attacks}, 
      author={Rongchang Li and Minjie Chen and Chang Hu and Han Chen and Wenpeng Xing and Meng Han},
      year={2024},
      eprint={2409.19521},
      archivePrefix={arXiv},
      primaryClass={cs.CR},
      url={https://arxiv.org/abs/2409.19521}, 
}

@misc{openai_gpt41,
  author       = {{OpenAI}},
  title        = {GPT-4.1 and GPT-4.1-mini},
  year         = {2024},
  howpublished = {\url{https://platform.openai.com/docs/models}},
  note         = {Accessed: 2025-03}
}

@misc{mistral_ministral3b,
  author       = {{Mistral AI}},
  title        = {Ministral-3B},
  year         = {2024},
  howpublished = {\url{https://docs.mistral.ai}},
  note         = {Accessed: 2025-03}
}

@inproceedings{lewis2020rag,
  title     = {Retrieval-augmented generation for knowledge-intensive NLP tasks},
  author    = {Lewis, Patrick and Perez, Ethan and Piktus, Aleksandra and others},
  booktitle = {NeurIPS},
  year      = {2020}
}

@misc{gao2024retrievalaugmentedgenerationlargelanguage,
      title={Retrieval-Augmented Generation for Large Language Models: A Survey}, 
      author={Yunfan Gao and Yun Xiong and Xinyu Gao and Kangxiang Jia and Jinliu Pan and Yuxi Bi and Yi Dai and Jiawei Sun and Meng Wang and Haofen Wang},
      year={2024},
      eprint={2312.10997},
      archivePrefix={arXiv},
      primaryClass={cs.CL},
      url={https://arxiv.org/abs/2312.10997}, 
}

@inproceedings{
yao2023react,
title={ReAct: Synergizing Reasoning and Acting in Language Models},
author={Shunyu Yao and Jeffrey Zhao and Dian Yu and Nan Du and Izhak Shafran and Karthik R Narasimhan and Yuan Cao},
booktitle={The Eleventh International Conference on Learning Representations },
year={2023},
url={https://openreview.net/forum?id=WE_vluYUL-X}
}

@misc{liu2024promptinjectionattackllmintegrated,
      title={Prompt Injection attack against LLM-integrated Applications}, 
      author={Yi Liu and Gelei Deng and Yuekang Li and Kailong Wang and Zihao Wang and Xiaofeng Wang and Tianwei Zhang and Yepang Liu and Haoyu Wang and Yan Zheng and Yang Liu},
      year={2024},
      eprint={2306.05499},
      archivePrefix={arXiv},
      primaryClass={cs.CR},
      url={https://arxiv.org/abs/2306.05499}, 
}

@inproceedings{10.5555/3666122.3669630,
author = {Wei, Alexander and Haghtalab, Nika and Steinhardt, Jacob},
title = {Jailbroken: how does LLM safety training fail?},
year = {2023},
publisher = {Curran Associates Inc.},
address = {Red Hook, NY, USA},
booktitle = {Proceedings of the 37th International Conference on Neural Information Processing Systems},
articleno = {3508},
numpages = {32},
location = {New Orleans, LA, USA},
series = {NIPS '23}
}

@INPROCEEDINGS{7917637,
  author={Sojka-Piotrowska, Anna and Langendoerfer, Peter},
  booktitle={2017 IEEE International Conference on Pervasive Computing and Communications Workshops (PerCom Workshops)}, 
  title={Shortening the security parameters in lightweight WSN applications for IoT - lessons learned}, 
  year={2017},
  volume={},
  number={},
  pages={636-641},
  doi={10.1109/PERCOMW.2017.7917637}}

@techreport{oecd_sme_digitalisation_2021,
  title       = {The Digital Transformation of SMEs},
  author      = {{Organisation for Economic Co-operation and Development}},
  institution = {OECD},
  year        = {2021},
  url         = {https://www.oecd.org/industry/smes/SME-Digitalisation-Policy-Perspectives.pdf},
  note        = {OECD Policy Perspectives}
}

@online{k3s_lightweight_kubernetes,
  title        = {K3s: Lightweight Kubernetes},
  author       = {{Rancher Labs}},
  year         = {2023},
  url          = {https://k3s.io/},
  urldate      = {2025-03}
}

@misc{australian_privacy_act_1988,
  title        = {Privacy Act 1988},
  author       = {{Australian Government}},
  year         = {1988},
  url          = {https://www.legislation.gov.au/Details/C2023C00197},
  note         = {Office of the Australian Information Commissioner (OAIC), accessed 2025-03}
}

@misc{australian_privacy_principles,
  title        = {Australian Privacy Principles},
  author       = {{Office of the Australian Information Commissioner}},
  year         = {2023},
  url          = {https://www.oaic.gov.au/privacy/australian-privacy-principles},
  note         = {Australian Government, accessed 2025-03}
}
\end{document}